\documentclass[aps,pre,reprint,superscriptaddress,amsmath,amssymb]{revtex4-1}

\usepackage{graphicx}
\usepackage{dcolumn}
\usepackage{bm}
\usepackage{hyperref}
\usepackage{comment}
\usepackage{color}
\usepackage{soul}
\begin{document}


\title{Measuring and upscaling micromechanical interactions in a cohesive granular material}
\author{Arnaud Hemmerle$^\ddagger$}
\thanks{These authors contributed equally to this work.}
\affiliation{Max Planck Institute for Dynamics and Self-Organization - Am Fassberg 17, 37077 G\"ottingen, Germany}
\affiliation{Synchrotron SOLEIL, L'Orme des Merisiers, Saint-Aubin, BP 48, 91192 Gif-sur-Yvette Cedex, France}
\author{Yuta Yamaguchi$^\ddagger$}
\thanks{These authors contributed equally to this work.}
\affiliation{Department of Earth and Planetary Science, University of Tokyo - 7-3-1 Hongo, Bunkyo, 113-0033 Tokyo, Japan}
\affiliation{Department of Earth and Space Science, Osaka University - 1-1 Machikaneyamacho, Toyonaka, 560-0043 Osaka, Japan}
\author{Marcin Makowski}
\affiliation{Max Planck Institute for Dynamics and Self-Organization - Am Fassberg 17, 37077 G\"ottingen, Germany}
\author{Oliver B\"aumchen}
\affiliation{Max Planck Institute for Dynamics and Self-Organization - Am Fassberg 17, 37077 G\"ottingen, Germany}
\affiliation{Experimental Physics V, Univ. of Bayreuth, Universit\"atsstr. 30, D-95447 Bayreuth, Germany}
\author{Lucas Goehring}
\email{lucas.goehring@ntu.ac.uk}
\affiliation{School of Science and Technology, Nottingham Trent University, Nottingham NG11 8NS, UK.}

\begin{abstract}
The mechanical properties of a disordered heterogeneous medium depend, in general, on a complex interplay between multiple length scales. Connecting local interactions to macroscopic observables, such as stiffness or fracture, is thus challenging in this type of material. Here, we study the properties of a cohesive granular material composed of glass beads held together by soft polymer bridges. We characterise the mechanical response of single bridges under traction and shear, using a setup based on the deflection of flexible micropipettes. These measurements, along with information from X-ray microtomograms of the granular packings, then inform large-scale discrete element model (DEM) simulations.  Although simple, these simulations are constrained in every way by empirical measurement and accurately predict mechanical responses of the aggregates, including details on their compressive failure, and how the material's stiffness depends on the stiffness and geometry of its parts.  By demonstrating how to accurately relate microscopic information to macroscopic properties, these results provide new perspectives for predicting the behaviour of complex disordered materials, such as porous rock, snow, or foam.
\end{abstract}
\maketitle




\section{Introduction}

Disordered porous materials are ubiquitous in nature and industry. The mechanical behaviour of these materials depends on the local interactions between their components, \cite{dvorkin1991, cubuk2017, harrington2018, jung2021} which can be of various shapes, compositions, and length scales, from nanometres in colloidal glasses to millimetres in sedimentary rocks.  Understanding and predicting the relationship between their microscopic structure and macroscopic mechanical response is of importance for applications in engineering, geoscience and materials science, and is the topic of active research in both experimental and numerical domains.\cite{kun2013, kun2014, neveu2016, cubuk2017, gaume2017, harrington2018, yamaguchi2018, ozawa2018,  wang2019, jung2021}  As an original example, it has been shown how birds such as swallows build strong nests using their saliva to form cohesive bonds between mud granules, in an analogy with the mechanical properties observed in artificial cohesive granular materials.\cite{jung2021}

Linking the mechanical properties of a disordered solid to its local structure presents difficulties. For example, the failure of a disordered material can suddenly change from ductile to brittle, depending on the density, strength, and plasticity of its contacts.\cite{shekhawat2013, driscoll2016, berthier2019, yamaguchi2020} Experimentally, it is hard to study such effects in realistic systems: these are complex by nature and have intrinsic characteristics that cannot be systematically varied. While different bottom-up approaches using model systems have emerged within the last ten years,\cite{delenne2009, arif2012, li2015, hemmerle2016a, schmeink2017} they are still rare. Thus, most progress in this field has been made in numerical modelling, \cite{jing2003, kun2013, kun2014, neveu2016, gaume2017, yamaguchi2018} with the inherent difficulty that a too simplistic model might not capture the complexity observed experimentally, while an over-detailed one may not give any useful general law.

\begin{figure}[t]
 \centering
 \includegraphics[width=1\linewidth]{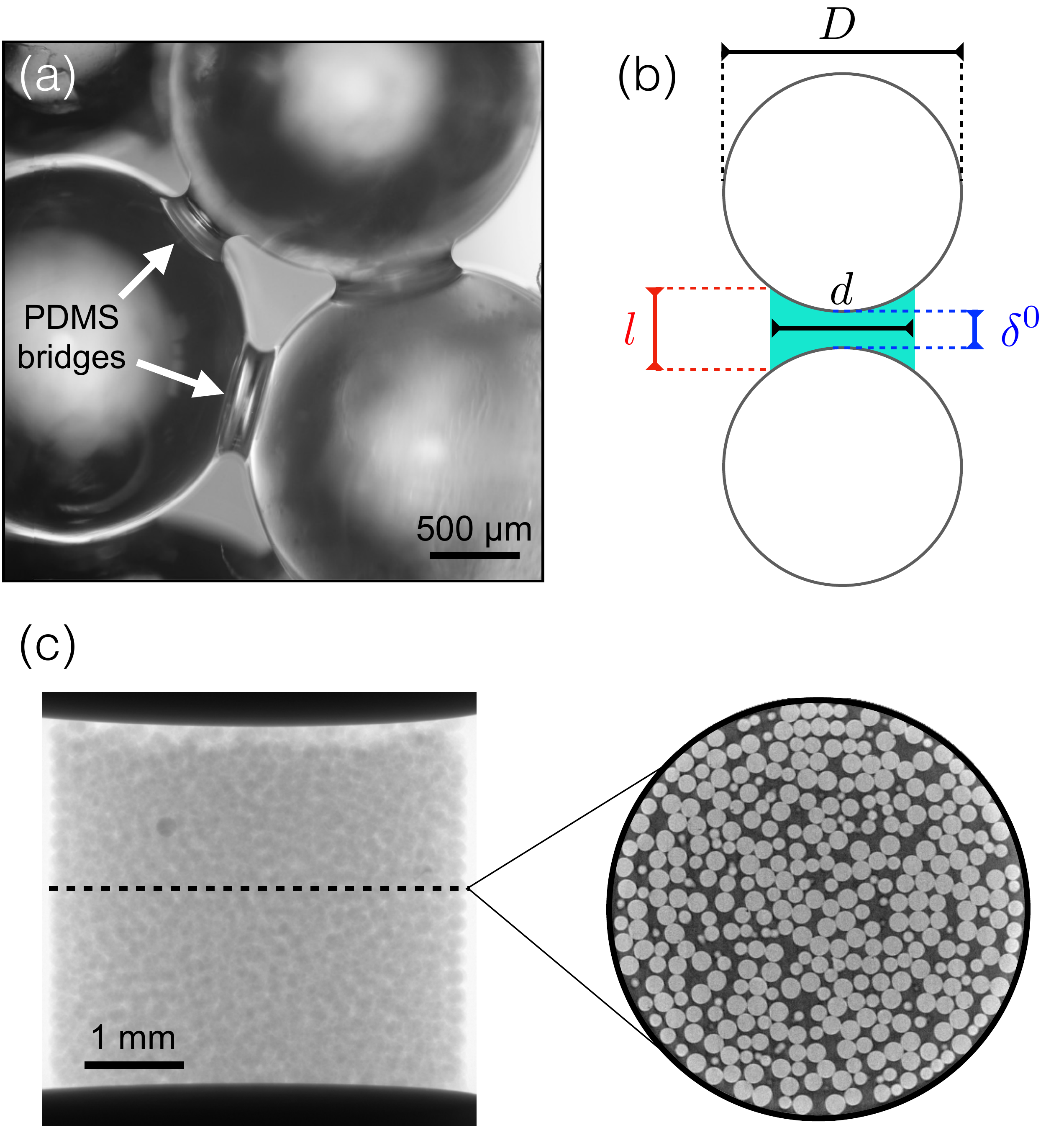}
 \caption{Structure of a cohesive granular material.  \textbf{(a)} A micrograph of a cohesive granular sample shows the capillary bridges of solid PDMS formed between glass beads. \textbf{(b)} We model each pair of connected beads as spheres of diameter $D$ linked by a truncated cylinder of diameter $d$ and height $l$ and separated by a gap of size $\delta^0$. \textbf{(c)} X-ray tomogram of a sample compressed {\it in situ} between two pistons. The inset shows a cross-section of the 3D data. The sample is a cylinder of 4.25 mm diameter $\times$ 4.05 mm height, made of beads of diameter $D =$ 200.9 $\mu$m.}
 \label{fig:Figure_Model}
\end{figure}

A tunable experimental model of a cohesive granular material has been recently introduced,\cite{hemmerle2016a, schmeink2017} which we will use here to help overcome these challenges.  It consists of glass beads held together by elastic bridges of polydimethylsiloxane (PDMS), a solidified elastomer, whose Young's modulus can be easily varied by changing its composition (see Fig.~\ref{fig:Figure_Model}). Despite its apparent simplicity this material shows a rich range of behaviour, allowing investigation of the mechanical properties of cohesive granulates in particular and of disordered solids in general.  Its elastic\cite{hemmerle2016a} and fracture\cite{schmeink2017} properties can be finely controlled by varying the stiffness, size, and density of the PDMS bonds linking the beads. The wide range of tunable parameters makes this clean and well-defined system an appropriate candidate to investigate scaling laws in the mechanics of disordered structures including, for examples, avalanches in snow, onsets of earthquakes, fracture of rocks and building materials, or poroelasticity and fluid-rock interactions in rocks under large hydrostatic pressure.

This cohesive granular material also shows intriguing mechanical properties. For instance, it fails under uniaxial compression by sliding along a shear plane (as expected for this type of material) but only after a remarkably large yield strain.\cite{hemmerle2016a} Such a feature is the result of an interplay between microscopic properties at the scale of individual beads and the large-scale (re-)organisation of the packing under mechanical load. A full understanding of the macroscopic properties of the material requires a good characterisation of the local contacts and a proper {\it upscaling} of microscopic laws towards larger assemblies of beads.
 
To this end, we report here direct measurements of the constitutive relationship of the bonds between individual particles in this type of cohesive granular material, and then demonstrate how to use these measurements to accurately predict the elastic properties of such a material.  To link the microscopic and macroscopic length scales we use a discrete element model (DEM), developed by some of us to investigate the failure mechanisms of cohesive granulates, and which incorporates a Griffith-like energy-based failure criterion for the bonds.\cite{yamaguchi2020}   Both the experiments and model involve hard particles, connected by softer bridges, and in this limit we show that the bridge geometry is a key parameter in upscaling mechanical properties.  The experimental observations will inform the parameters used in the model and we will then show that this approach can reproduce, both qualitatively and quantitatively, key features observed in the experiments, such as the results of uniaxial compression tests.  

\section{Micromechanics of cohesive granulates}
\label{micro}

We first characterised the micromechanical interactions between particles in a cohesive granular medium.   If two particles or grains are mechanically bound to each other, they may interact \textit{via} normal and tangential displacements, and by rolling and twisting. These mechanisms can be modelled by elastic springs with associated spring constants for the normal and tangential deformations ($k_n^{\rm bond}$ and $k_t^{\rm bond}$, respectively), and bending stiffnesses for the rolling and twisting ones. Aiming for a micromechanical model which reproduces the essential properties of this type of material with as few parameters as necessary, we developed a protocol for measuring the normal and tangential spring constants over a single pair of beads, while assuming contributions from bending and twisting to be less relevant. Here, we describe these methods and give direct observations of the elasticity of the bonds between particles taken from a cohesive granular material. 

\begin{figure*}[t!]
\centering
  \includegraphics[width=0.8\textwidth]{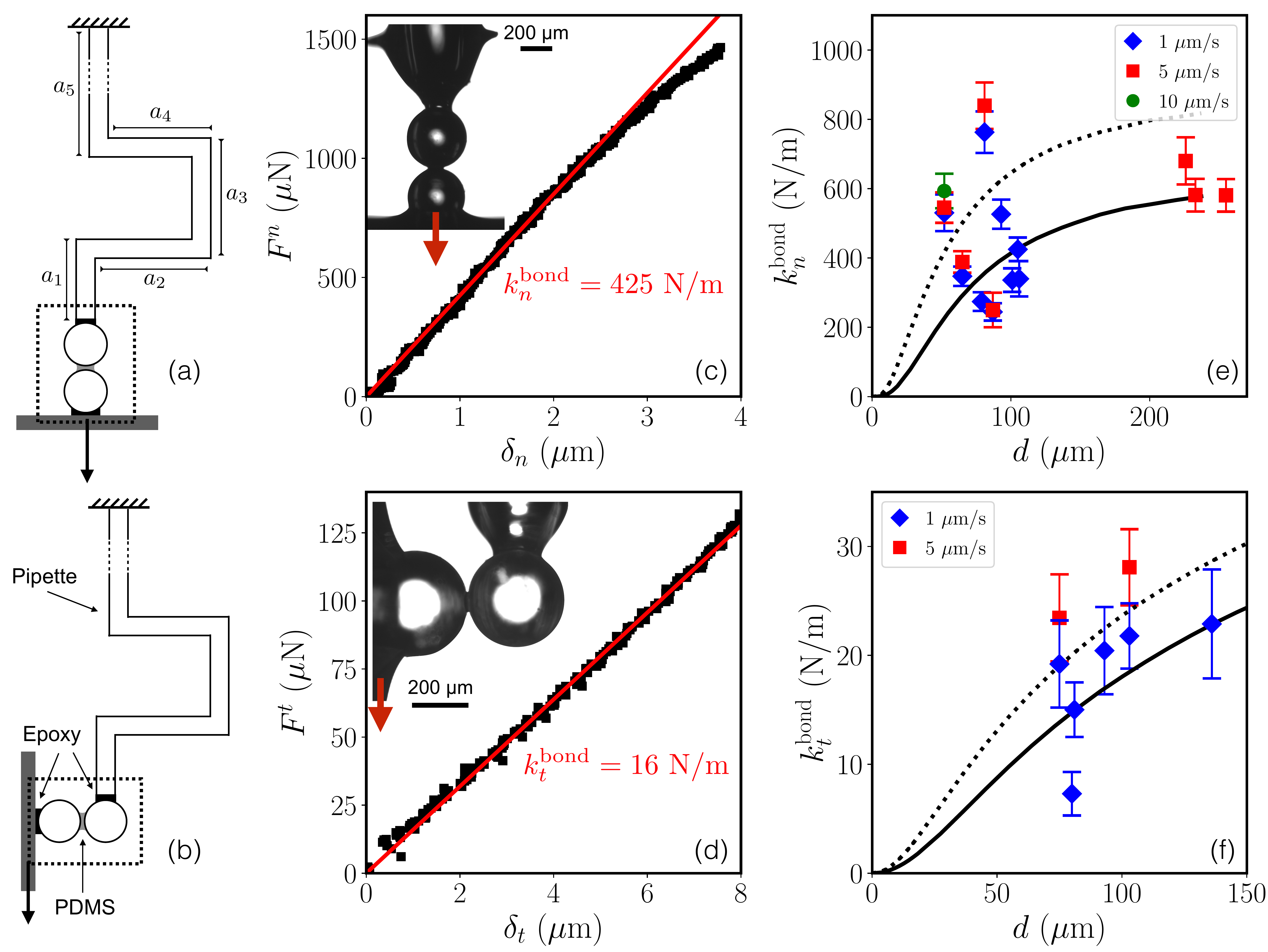}
  \caption{Schematics of the micromechanical tests in the normal \textbf{(a)} and tangential \textbf{(b)} configurations. The dotted frames correspond to the images shown as inserts in the example force-displacement measurements given in \textbf{(c)} and \textbf{(d)}. Linear fits (red lines) to the force-displacement data give the spring constants $k_n^{\rm bond}$ or $k_t^{\rm bond}$. Panels \textbf{(e,f)} summarise the spring constant measurements made of different bridge diameters $d$ and at various speeds. The lines are the results of simulations of bridges in similar conditions, assuming an initial bridge length, $\delta^0$, of either 1~$\mu$m (dotted line) or 2~$\mu$m (solid line).}
  \label{fig:Figure_pipette_schem}
\end{figure*}

\subsection{Experimental methods}

Micromechanical testing was conducted using flexible glass micropipettes (see Fig.~\ref{fig:Figure_pipette_schem}), in an approach adapted from one used to measure the forces exerted by cells and other micro-organisms.\cite{backholm2019} 

Samples of cohesive granular material were prepared, following published methods.\cite{hemmerle2016a} The elastic bridges consist of polydimethylsiloxane (PDMS, Sylgard 184, Dow Corning), a curable elastomer, using a mass ratio of base to cross-linker of 40:1.  Uncured PDMS was mixed with glass beads of 365 $\mu$m diameter (Sigmund Lindner GmbH, 5$\%$ polydispersity) and at a polymer volume fraction $W=2.3\%$, poured into a cylindrical mould and cured at 90$^{\circ}$C overnight. The Young's modulus of the PDMS itself, $E_p=50\pm 5$ kPa, was measured by unconfined uniaxial compression of a cylinder of cured polymer. The high base to cross-linker ratio, and low $E_p$, allows the bridges to be soft enough to give measurable deformations during the micromechanical tests, while still behaving rigidly and elastically. 

For each test a single pair of beads, connected by a PDMS bridge, was detached from a sample of cohesive granular material using a scalpel under an inverted microscope (IX-73, Olympus), taking care not to stretch or bend the bridge. The beads were moved onto a glass slide, allowing for half of one bead to protrude over the edge of the slide. A droplet of epoxy was deposited onto the end of the dangling bead, which was then quickly put into contact with a flat silicon wafer. After the epoxy was cured the glass slide was removed, leaving the pair of beads attached normally to the wafer. Finally, a glass micropipette was glued onto the other bead, resulting in configurations as shown in Fig.~\ref{fig:Figure_pipette_schem}. 

The micropipettes are narrow glass filaments (borosilicate glass capillaries, World Precision Instruments, TW100-6), pulled by an horizontal pipette puller (P-97 Micropipette Puller, Shutter Instruments) and bent into a characteristic U-shape in a microforge (MF-900, Narishige). This shape gives them an elastic response in extension, as the glass would break before any plastic deformation would occur, and allows pure normal ($F^{n}$) or shear ($F^{t}$) forces to be directed across the bridge, minimising any off-axis contributions. Their spring constant is fixed by their geometry and measured using a pre-calibrated AFM cantilever (Veeco).\cite{backholm2019} For example, one typical pipette had a diameter of 0.3 mm and side lengths, as defined in Fig.~\ref{fig:Figure_pipette_schem}, of $a_1=2.5$ mm, $a_2=13$ mm, $a_3=3.5$ mm, $a_4=8.5$ mm and $a_5=117$ mm, resulting in a spring constant of $40\pm 2$ N/m. During an experiment, one end (top of $a_5$) of the pipette was kept fixed, while the wafer was mounted on and moved by a motorised translation stage (Newport, Conex LTA-HS). 

Images were recorded of experiments at 10 frames/second with a digital camera (FLIR Systems, Grasshopper 3, GS3-U3-41C6M-C) and a 4$\times$ (1.37 $\mu$m per pixel) or 10$\times$ objective (0.55 $\mu$m per pixel). Bead positions were measured using cross-correlation image analysis with a sub-pixel resolution.\cite{backholm2019} Glass spikes made on the pipette with the microforge allowed for independent tracking of the beads and pipette.  The normal/extensional ($\delta_n$) or tangential ($\delta_t$) deformation of the bridge was deduced from the relative displacement of the two beads, while the force exerted on the pipette was calculated from its deflection and spring constant.\cite{backholm2019}  Finally, the bead diameters were found by fitting circles to their shapes in the microscopy images, while the bridge diameters, $d$, were measured across their middle. For both diameters, the measurement uncertainty is estimated to be $\pm$ 2 $\mu$m. The initial bridge spanning distance, $\delta^0$, \textit{i.e.} the initial surface-to-surface separation between the beads, was always found to be lower than $2$ $\mu$m, but could not be measured more accurately.

\subsection{Results of micromechanical tests}

Tests were performed on bridges with diameters $d$ between $50$ and $270$ $\mu$m (see Movies S1 and S2 in the ESI for examples of normal and shear tests, respectively). In each case a spring constant was calculated from a linear fit to the force-displacement curve in the linear regime of deformation, as demonstrated in Fig.~\ref{fig:Figure_pipette_schem}(c,d). The restoring forces were linear over a wide range of displacements, especially compared to the initial particle separation, $\delta^0$. This confirms that we can treat the bonds as Hookean springs.  We further checked that the response was in the quasi-static limit by repeatedly testing some bridges at increasing deformation speeds of 1 to 10 $\mu$m/s, with minimal differences in the spring constants. 

As shown in Fig.~\ref{fig:Figure_pipette_schem}(e,f), values for the normal spring constants, $k^{\rm {bond}}_n$, lay in the range of $200-800$ N/m, while tangentially, $k^{\rm {bond}}_t$ was between $5-30$ N/m.  Despite the dispersion of results for bridges prepared in a similar manner, it is clear that $k^{\rm {bond}}_n\gg k^{\rm {bond}}_t$, meaning that bridges are strong in tension, but easy to shear. 

\subsection{Scaling spring constants: FEM simulation}
\label{COMSOL}

The micromechanical measurements were made for beads of one particular size, $D$, and a fixed polymer content $W$ and stiffness $E_p$.   In order to explore how a bridge's spring constants depend on these variables, we performed simulations replicating our micromechanical tests, but for various bond geometries.

To this end, we used COMSOL Multiphysics to build a finite-element model (FEM) simulation of an elastic bridge. The bridge was modelled as a cylinder of diameter $d$, truncated by two spherical caps of diameter $D$ and surface separation $\delta^0$, as in Fig.~\ref{fig:Figure_Model}(b).  It was treated as a linear elastic material, with Young's modulus $E_p$ and Poisson ratio $\nu = 0.49$.  A normal or tangential displacement was imposed on one of the sphere-bridge interfaces, while the other was kept fixed, using no-slip boundary conditions for both.  The total reaction force exerted on either interface was measured as a function of the displacement, from which follow the spring constants $k^{\rm {bond}}_n$ and $k^{\rm {bond}}_t$ of the simulated bridge.

The key result is shown in Fig.~\ref{fig:Figure_pipette_schem}(e,f). Namely, if we use the experimental values of $D$, $d$ and $E_p$, and take $\delta^0=2$ $\mu$m, then the simulations accurately predict the magnitudes of $k^{\rm {bond}}_n$ and $k^{\rm {bond}}_t$ and their relatively weak dependence on $d$.  The simulation's sensitivity to $\delta^0$ can also potentially explain why there is such a large variation in the experimentally measured spring constants.  For example, by adjusting $\delta^0$ between 1 and 2  $\mu$m (which is below our experimental resolution), the simulated spring constants span the range of most experimental values.  

Otherwise, the scaling of the spring constants with the bridge geometry is not trivial, although it is clear that there should be a linear dependence on $E_p$.  In the following we will therefore use the COMSOL simulations to predict representative values for $k_n^{\rm bond}$ and $k_t^{\rm bond}$, given specific bead and bridge sizes.  In particular, we are interested in mimicking experimental uniaxial compression tests with a discrete element model of large-scale bead assemblies, and for these cases have simulated bridges with microscopic parameters matching experimental observations (see Table~\ref{tab:parameter_info}). 

\begin{table*}[t!]
    \centering
    \setlength{\tabcolsep}{5pt}
    \begin{tabular}{|l|c|c|c|c|}\hline
         Property & Symbol & Micromechanical tests & Sample A& Sample B
           \\ \hline 
    	Bond Young's modulus & $E_p$ & $50\ {\rm kPa}$ & $1.5\ {\rm MPa}$ & $0.64\ {\rm MPa}$ \\ 
	Bond diameter             & $d$      & $50-270\ {\rm \mu m}$  & $73.9\ {\rm \mu m}$ &  $72.1\ {\rm \mu m}$  \\ 
	Bead diameter             & $D$      & $365\ {\rm \mu m}$ &  $200.9\ {\rm \mu m}$ & $200.9\ {\rm \mu m}$\\ 
	\hline
        Bond stiffness, normal    & $k_n^{\rm bond}$ &$200-800\ {\rm N/m}$ &  5760\ {\rm N/m} &  2345\ {\rm N/m}\\
        Bond stiffness, tangential  & $k_t^{\rm bond}$ & $5-30\ {\rm N/m}$ &325 \ {\rm N/m}& 131  \ {\rm N/m}\\ 
         \hline     
    \end{tabular}
    \caption{Key parameters used in the micromechanical modelling.  Values are given for the materials used in the micromechanical tests (Section~\ref{micro}), as well as the two samples, A and B, for which the initial bead positions are known from X-ray microtomography (Section~\ref{microtomography}).   In all cases $E_p$, $d$ and $D$ are measured experimentally.  The bond stiffnesses are directly measured for the micromechanical test samples, and these results used to validate a FEM simulation of an elastic bridge, which is then used to predict $k_n^{\rm bond}$ and $k_t^{\rm bond}$ for the other samples.}
    \label{tab:parameter_info}
\end{table*}

\section{Upscaling micromechanical interactions}
\label{Numerics}

Cohesive granular materials can behave as linear elastic solids, with well-defined failure conditions.\cite{hemmerle2016a, schmeink2017, cubuk2017, gaume2017, delenne2009, jiang2013, wang2019, brendel2011} However, predicting the homogeneous elastic response of members of this class of material from knowledge of their micro-structure, {\it i.e.}\,upscaling a microscopic model, is made difficult by their discrete and disordered nature. Even more challenging is the prediction of how, and when, such a system will fail, because of the importance of nonlinear effects such as force chains and strain localisation when approaching failure.\cite{gilabert2008,brendel2011,baud2014,roy2016,gaume2017, mcbeck2019,yamaguchi2020}

Here, we pursue this goal of upscaling the micromechanical properties of a cohesive granular material.   Experimentally, we imaged such materials using X-ray microtomography, and extracted the positions of the constituent beads along with statistical information about the bridge networks. These measurements, combined with the micromechanical characterisation of single bridges (see Section 2), are then used as inputs for a minimal model of cohesive granular materials.\cite{yamaguchi2020}  This model is thus constrained in all its parameters by observations at the microscopic scale.   In Section 4, we will then simulate unconfined uniaxial testing of the modelled material, and find that the results compare favourably to {\it in situ} experimental compression tests. 

\subsection{X-ray microtomography: Bead and bridge geometry}
\label{microtomography}

X-ray microtomography data obtained in a previous study\cite{hemmerle2016a} were further analysed in the present work (sample A, see Fig.~\ref{fig:Figure_Model}(c)), and complemented by data from a similar sample which featured softer bonds (sample B). Briefly, unconfined uniaxial compression tests of cohesive granular samples were performed {\it in situ} within an X-ray micro-computed tomography system (GE Nanotom).  The samples were composed of monodisperse beads of diameter $D = 200.9 \pm 1.9 \ \mu$m  (Whitehouse Scientific), mixed  with PDMS of Young's modulus $E_p=1.5\pm 0.15$ MPa (sample A) or $E_p=0.64 \pm 0.05$ MPa (sample B).  These materials were shaped into cylinders of 4.25 mm diameter $\times$ 4.05 mm height for sample A, and 4.25 mm diameter $\times$ 4.82 mm height for sample B.  Tomography scans were performed before compression, and consisted of sets of 2000 projections with a resolution of 1132 $\times$ 1132 pixels and a voxel size of 5 $\mu$m.  The samples were then compressed at a speed of $5 \ \mu$m/min, and their stress-strain curves will be compared with simulations in Section 4, and Fig.~\ref{fig:Figure_comparison}.  

Beads were detected within the reconstructed sample volumes using Matlab, and particle positions determined by finding the centroid of each individual connected volume after segmentation, with a precision of about one voxel (see Tables S1 and S2 in the ESI for particle positions and Movies S3 and S4 for illustrating their detection).  However, the X-ray contrast between the PDMS bridges and the glass beads was insufficient for the automatic detection of which particles were linked by bridges.  Instead, we manually counted bridges on beads within the interior of the sample.  For example, in sample B the coordination number $Z$, here defined as the average number of bridges per bead, was found to be $7.4\pm 0.1$ (standard error), which is consistent with previous results from a similar sample.\cite{schmeink2017}  The distributions of bridge diameters were also extracted by manual inspection of the microtomograms.  For example, for sample A measurements from 100 bridges were well-approximated by a Gaussian distribution with a mean of $73.9\ {\rm \mu m}$ and standard deviation of 12.7 $\mu$m. 

The characteristic properties of the two experimental samples are given in Table~\ref{tab:parameter_info}, along with the bond stiffnesses estimated through the micromechanical testing and COMSOL bridge model.  

\subsection{Discrete element model of cohesive granulates}

To link the microscopic measurements to the bulk properties of a cohesive granular material, we used a discrete element model (DEM) that simulates the interactions between particles and their bonds.  As the development of this model has been detailed elsewhere,\cite{yamaguchi2020} we focus only on its main features here, including the modifications made for this study.  

\subsubsection{Model setup and compression test}
\label{Preparation}


\noindent{\bf{Bead Positions.}} The initial bead locations are taken from the relative positions of particles as measured in either of the two microtomograms.  The natural dispersion in bead sizes and voxel resolution of the tomography data leads to a small uncertainty in these positions, which could result in slightly overlapping beads. We avoided this issue by randomly varying the diameter of each particle with the Box-Muller method~\cite{box1958} using a Gaussian distribution with a mean diameter of $D=200.9$ $\mu$m and a standard deviation of $5\%$, while constraining the sizes of the particles to prevent overlap.  Ensembles of at least 5 different realisations of particle sizes were used, and the error bars in Figs.\ \ref{fig:Figure_E_vs_Ep} and \ref{fig:Figure_E_vs_W} give the standard deviations of measurements on such ensembles. 

\noindent{\bf{Bridge network.}} To match the experimental value of the average coordination number (\textit{i.e.} bonds per particle), nearby particles were considered to share a bond at the start of the simulation only if their initial surface-to-surface distance was smaller than $0.064\,D$; this condition leads to $Z=7.4$ for particles in the interior of the simulated packing. Aiming for a model with as few parameters as possible, we assign all bridges in each sample the same values of $k^{\rm {bond}}_n$ and $k^{\rm {bond}}_t$, estimated from the average diameters $D$ and $d$ and using $\delta^0=2 \, \mu$m, unless otherwise mentioned.  In other words, for the purposes of setting up the bond stiffnesses we treat all bonds as identical, rather than introducing a function that would depend on their specific initial geometries.  Spring constants corresponding to samples A and B are given in Table~\ref{tab:parameter_info}.  These were evaluated by a look-up-table of discrete values from the FEM calculations, giving a precision of about 2\%.


\noindent{\bf{Compression test.}} Once a model sample has been configured it is confined between two flat, rigid walls. Uniaxial compression tests are then simulated by keeping the top wall fixed and moving the lower one upwards at a constant speed.  For example, sample A was compressed at $46.4\ {\rm mm/s}$, corresponding to a speed of 10$^{-4}$ in the non-dimensionalised terms of the simulation,\cite{yamaguchi2020} and the same dimensionless speed was used for all simulations.  This velocity is a compromise between requiring the dynamics to be slow enough to reproduce the quasi-static experiments, while keeping a reasonable computation time.   

\subsubsection{Model for the interactions}
 The DEM simulates the dynamics of a collection of spherical particles that can interact with each other through contact forces (glass-glass) and cohesive bonds (PDMS), modelled as linear, Hookean springs. Each bead, $i$, is represented by a sphere with mass $m_i$, centre position $\bm{r}_i$,  diameter $D_i$, angular velocity $\bm{\omega}_i$ and moment of inertia $I_i$. The particle can interact with its neighbours through normal forces and torques.  The resulting translational and rotational equations of motion are
\begin{align}
    &m_i\frac{d^2\bm{r}_i}{dt^2}=\sum_{j\neq i} (F_{ij}^n \bm{n}_{ij} + \bm{F_{ij}}^t ),\label{eq:eqm}\\
    &I_i \frac{d\bm{\omega}_i}{dt} = \frac{D_i}{2} \sum_{j \neq i} \bm{n}_{ij}\times \bm{F_{ij}}^t,
    \label{eq:rotation}
\end{align}
where $\bm{n}_{ij}=(\bm{r}_i-\bm{r}_j)/|\bm{r}_i-\bm{r}_j|$ is the normal unit vector between particles $i$ and $j$ and where their interaction force is decomposed into a normal component, $F_{ij}^n$, and any tangential forces, $\bm{F}_{ij}^t$.   

The normal force between two particles, $F_{ij}^n$, has three possible contributions: a contact force due to particle overlap, $F_{ij}^c$, an elastic force due to a bond, $F_{ij}^{\rm bond}$, and a dissipation force, $F_{ij}^{\rm diss}$.  The first two of these are conservative, and expressed as
\begin{align}
    F_{ij}^c + F_{ij}^{\rm bond}
    = \begin{cases}
    	-k_n^{\rm glass}\delta_{ij}^n + k_n^{\rm bond} \delta_{ij}^0 & \delta_{ij}^n \le 0,\\
	- k_n^{\rm bond}\left( \delta_{ij}^n - \delta_{ij}^0\right) & \delta_{ij}^n > 0,
    \end{cases}\label{eq:contact_force}
\end{align}
where $\delta_{ij}^n = |\bm{r}_i-\bm{r}_j|-(D_i+D_j)/2$ is the surface separation of the particles.  The terms involving $k_n^{\rm bond}$ are only applied if there is a bond between the two particles.  For bead overlap we set the spring constant $k_n^{\rm glass}=10 k_n^{\rm bond}$, which is high enough to represent a system of stiff beads connected by softer springs, but low enough to avoid numerical instabilities; we have confirmed\cite{yamaguchi2020} that simulation results are robust to changes in $k_n^{\rm glass}$.  Finally, we include dissipation due to particle deformation as $F_{ij}^{\rm diss} =  -\zeta (\bm{v}_{ij}\cdot\bm{n}_{ij})$, where $\bm{v}_{ij}$ is the relative velocity between beads $i$ and $j$, and $\zeta$ is a dissipation rate.\cite{yamaguchi2020}   

Tangential forces can also arise between particles which share a bond, or which overlap.  For this, a linear restoring force is generated by any relative tangential displacement of two beads around their first point of contact (\textit{i.e.} initial positions, for a bond).  A cohesive bond is treated as a spring with a tangential spring constant $k_t^{\rm bond}$.  For particle overlap, as with normal forces, we use $k_t^{\rm glass} = 10 k_t^{\rm bond}$.  In either case the tangential force is limited by a Coulomb friction law with a friction coefficient $\mu=0.5$.\cite{penskiy2011} 

The interaction of a particle with either the top or bottom wall is treated the same way as with another particle, although no bonds need to be considered, only overlap.  The total compressive stress in the system, $\sigma$, is then the sum of all the forces acting on a wall, divided by the initial cross-sectional area of the sample.

Finally, under significant enough tension or shear, a PDMS bond will fail; fracture testing has shown that the dominant mode of failure is by bonds detaching, or peeling away from the glass beads. \cite{schmeink2017}  For the corresponding mechanism in our model we adopt a Griffith-like failure criterion, by comparing the strain and surface energies of a bond.  This leads to the failure condition\cite{yamaguchi2020}
\begin{eqnarray}
\label{criterion}
	\Delta r_n^2 + \frac{|\bm{\delta_{ij}^t}|^2}{12} &\ge& 2 \frac{Gl_{ij}}{E_p}.
	\label{energy_balance_2}
 \end{eqnarray}
For this, the bond is treated as a cylinder of height $l_{ij}$. Heterogeneity in the failure condition is provided through the distribution of values of $l_{ij}$.  The normal displacement of the bond is $\Delta r_n = \delta_{ij}^n - \delta_{ij}^0$, and $G = 7\ {\rm J/m^2}$ is the measured interfacial energy between PDMS and glass.\cite{chopin2011,schmeink2017} Whenever a bond meets this criterion, it is removed from the simulation.

\section{Elasticity of cohesive granulates}

Having measured the constitutive relationship of the individual bonds between the particles of a cohesive granular material, and outlined a model of this class of material that is designed around these microscopic measurements, we now test the predictions of this model.  Specifically, we will compare the stress-strain curves of simulations of uniaxial compression against experiments, and then explore how the elastic response changes with the stiffness and size of the bridges.  A related manuscript\cite{yamaguchi2020} makes use of the same model, but focuses on the macroscopic failure mechanisms, and how these depend on the packing density of the material.  

\subsection{Stress-strain curves}

\begin{figure}[t!]
 \centering
 \includegraphics[width=1\linewidth]{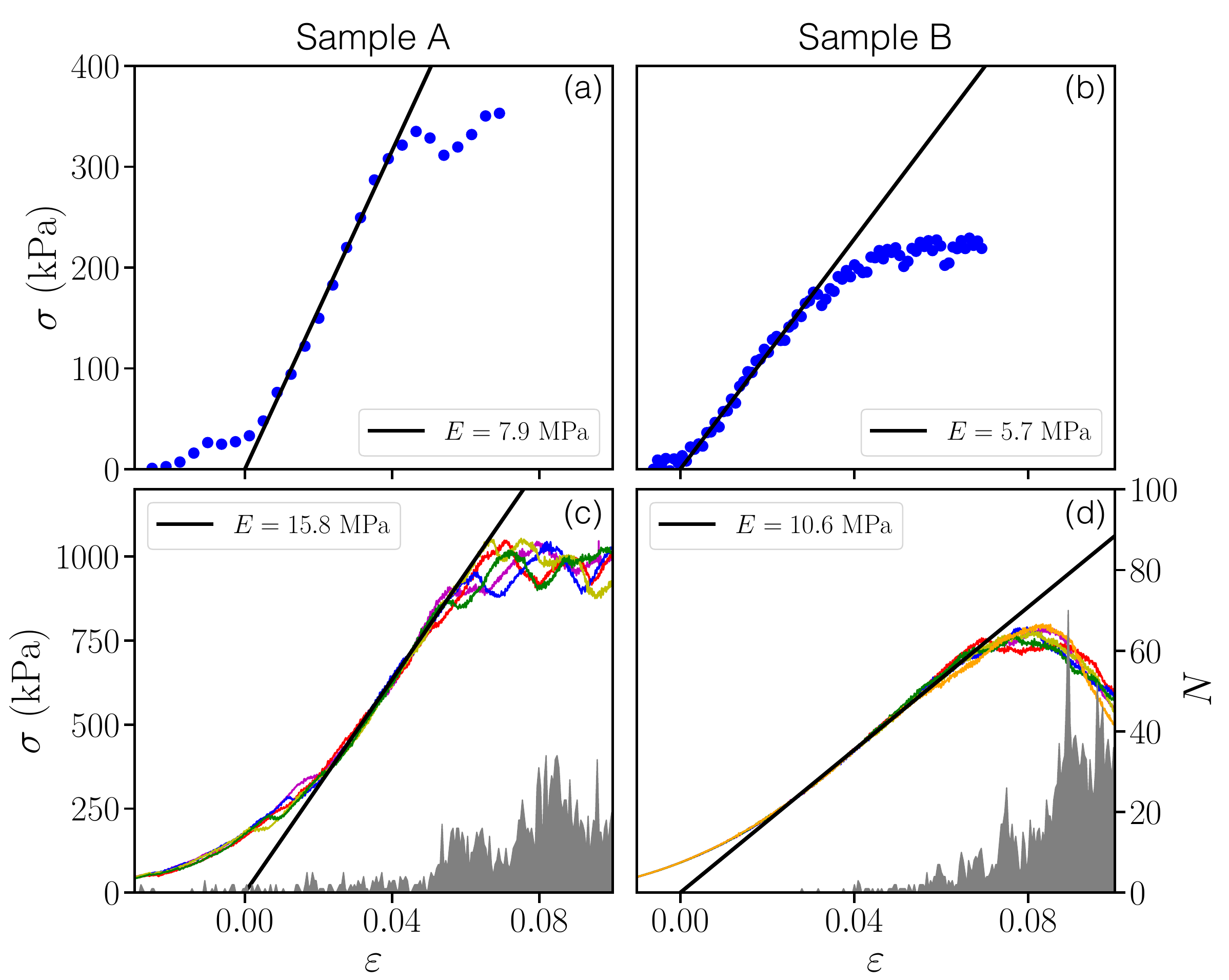}
 \caption{Comparison of experimental \textbf{(a,b)} and simulated \textbf{(c,d)} stress-strain curves. In each case the Young's modulus, $E$, is obtained by a least-squares fit to the linear region (black lines). Each simulated curve corresponds to a different initial realisation of the particle diameters, but using the same particle positions. Panels \textbf{(a,b)} share the same $y$-axis, as do \textbf{(c,d)}.  The grey area in \textbf{(c)} and \textbf{(d)} gives the number of bonds that break, $N$, in each increment of strain.}
 \label{fig:Figure_comparison}
\end{figure}

First, we compare the results of uniaxial compression in matched experimental and simulated geometries. The experiments are described in Section~\ref{microtomography}, with a typical sample shown in Fig.~\ref{fig:Figure_Model}(c), and the compression tests follow protocols described in more detail elsewhere.\cite{hemmerle2016a}  We show in Fig.~\ref{fig:Figure_comparison}(a,b) the normal stress, $\sigma$, measured as a function of the compressive strain, $\epsilon$, for samples A and B.  The corresponding results from simulations in the same geometries (\textit{i.e.} using values from Table~\ref{tab:parameter_info} and particle positions taken from X-ray tomograms) are given in Fig.~\ref{fig:Figure_comparison}(c,d). Videos of the simulations are given as Movies S5 and S6 in the ESI. 

For both samples the model accurately captures the shape of the stress-strain curves, including the yielding behaviour at large $\epsilon$, although the modelled system behaves about twice as stiffly as the experiments.  Different realisations of the simulations show little variation in the predicted response.   All curves start with a gradual build-up of stress, with a concave shape corresponding to the progressive contact between the rough surface of the sample and the piston.\cite{hemmerle2016a} A linear response follows, from which the Young's modulus is measured by a least-squares fit.  Here, we have used an extrapolation of the linear regime of deformation to define our zero-point of strain (\textit{i.e.} $\epsilon = 0$).   With this convention, the linear regime extends up to about 4\% in the experiments as well as in the simulations, after which point failure sets in.

In the simulations, we also tracked the number of bonds broken in each time step, binned into strain windows of size $\Delta\epsilon = 4.2\times 10^{-4}$.   These data, shown in  Fig.~\ref{fig:Figure_comparison}(c,d) for representative cases, give a measure of the microcrack activity, which remains low until the end of the linear regime.  This confirms that the response is essentially elastic until the peak stress or failure point is reached.  Our companion paper\cite{yamaguchi2020} explores the failure processes of cohesive granular materials in more detail; here, we simply note that failure proceeds by a shear band forming across the sample (see Movies S5 and S6 in the ESI), as in the experiments.\cite{hemmerle2016a}

\subsection{Variation with polymer stiffness}
\label{polymer}

The elastic responses of diverse types of cohesive granular materials are known to be controlled by the cohesive component (\textit{i.e.} the matrix), despite the fact that it represents only a small minority phase.\cite{delenne2009, hemmerle2016a, wang2019, jung2021}  This is particularly true when the matrix material is softer than the particles or grains it binds together.  For our samples the Young's modulus of the material, $E$, increases with, and is about an order of magnitude higher than that of, the stiffness of the PDMS composing its cohesive bridges, $E_p$.\cite{hemmerle2016a, schmeink2017} 

\begin{figure}[t!]
 \centering
 \includegraphics[width=.9\linewidth]{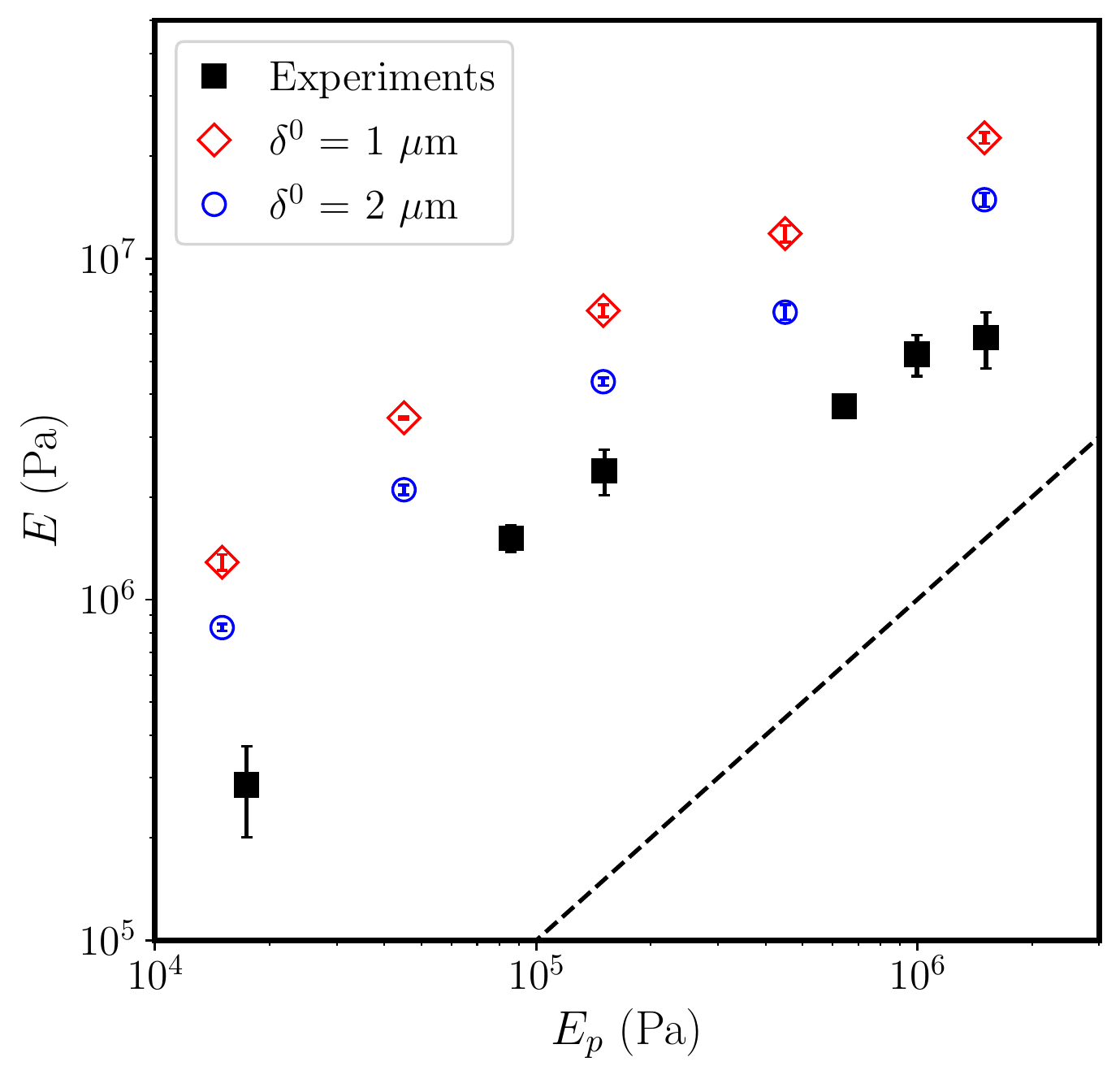}
 \caption{Comparison of how the Young's modulus of a cohesive granular sample, $E$, depends on the Young's modulus of the bridges composing it, $E_p$. The simulations (open symbols) reproduce the experimental relationship (closed squares) seen between $E$ and $E_p$, and show the importance of the microscopic bridge length $\delta^0$ in determining the macroscopic stiffness of the sample. The dashed line shows $E=E_p$, for comparison.
}
 \label{fig:Figure_E_vs_Ep}
\end{figure}

In order to compare our model predictions to the experimental results, we systematically varied $E_p$ in simulations of sample A, keeping the other parameters fixed. The results are shown in Fig.~\ref{fig:Figure_E_vs_Ep} alongside those from similar experiments\cite{hemmerle2016a} (note that these used slightly larger beads with higher polydispersity: $D=210$ $\mu$m $\pm$ 5 $\%$). The simulations correctly capture the scaling of $E$ with $E_p$, including the deviation from a linear dependency of $E\propto E_p$ for stiff aggregates, although they again predict stiffer samples than seen experimentally. Investigating this discrepancy in more detail, we repeated the simulations using spring constants calculated assuming initial particle separations of $\delta^0=1$ $\mu$m (rather than $2$ $\mu$m, see section 2.3).  The two different sets of spring constants, which span the range of behaviours observed by the micromechanical testing, had a significant impact on the stiffness of the aggregate, with $E$ changing by about 60\%. This increase is comparable to the increase in $k_n^{\rm bond}$, and demonstrates the sensitivity of the mechanical properties of this class of materials to the exact geometry and nature of the bonds between particles.  

\subsection{Variation with polymer content}

\begin{figure}[t!]
 \centering
 \includegraphics[width=.9\linewidth]{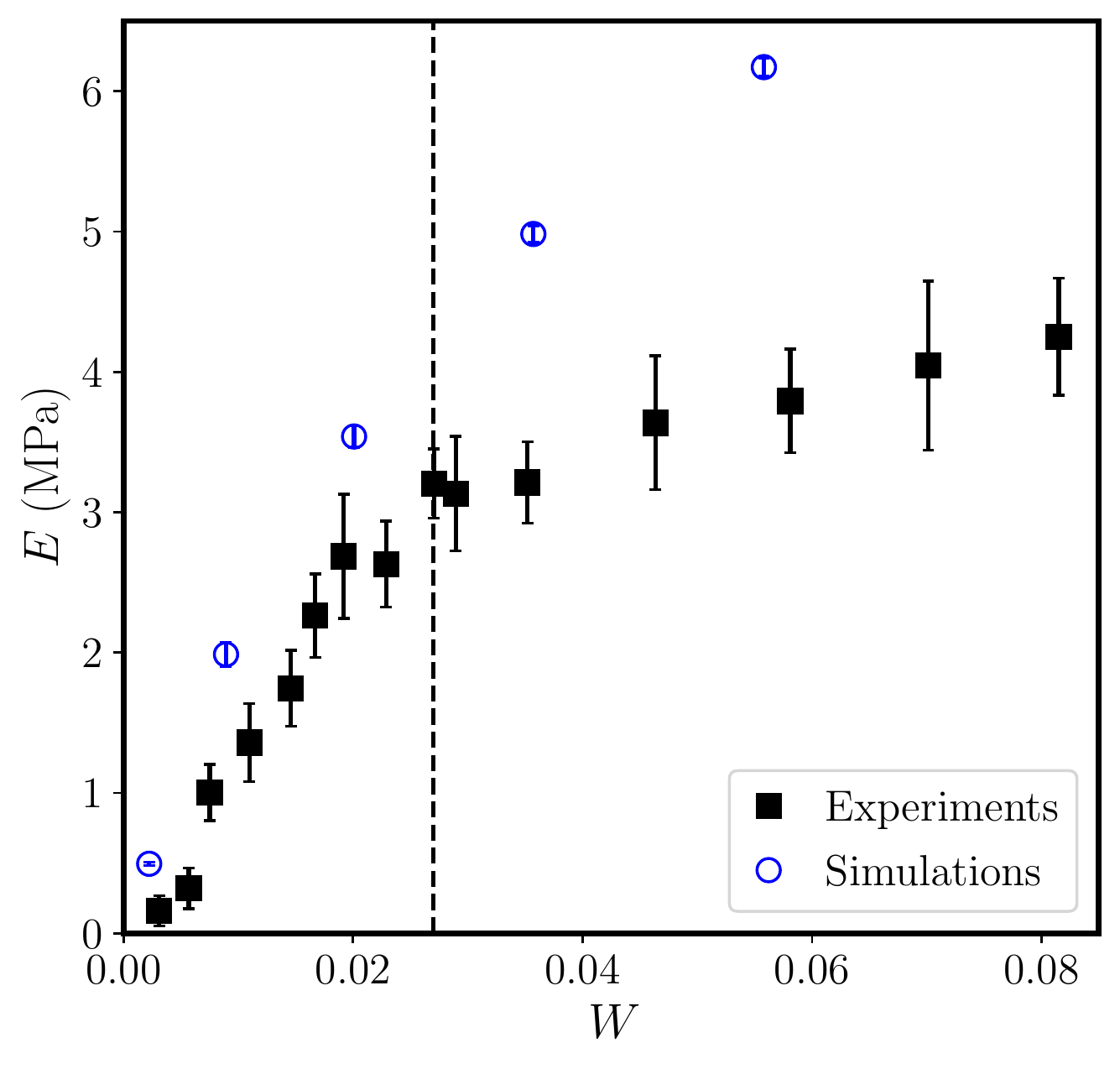}
 \caption{The Young's modulus of the material, $E$, increases with increasing content of PDMS, $W$, in a similar manner in both the experiments (closed squares) and the simulations (open circles), up to $W=2.7\%$ (dashed line). Experimentally, this value  corresponds to the appearance of trimer structures in the system,\cite{hemmerle2016a} while only pendular bridges are observed for lower $W$. Here, $E_p=250$ kPa in both experiments and simulations.
}
 \label{fig:Figure_E_vs_W}
\end{figure}

Finally, we focus on the stiffness of the aggregate when varying the volume fraction of polymer in the material, $W$. Experimentally, $E$ increases approximately linearly with $W$ up to a critical value of $W^\star = 2.7 \%$.  After this point a much slower increase of the stiffness with polymer content is observed (see Fig.~\ref{fig:Figure_E_vs_W}).\cite{hemmerle2016a} This change coincides with the point where the polymer bridges coalesce into more complex structures like trimers (\textit{i.e.} the pendular-funicular transition\cite{scheel2008,hemmerle2016a}).  In other words, above $W^\star$ the addition of more matrix material does not strengthen the bonds between particles, but rather fills in the larger pore spaces.

To test the model's response in the pendular or capillary regime, we systematically varied the bridge diameter $d$ in simulations of sample A, while fixing all other parameters.  The spring constants $k_n^{\rm bond}$ and $k_t^{\rm bond}$ were calculated as in Section 2.3, for each value of $d$.  Empirically,\cite{hemmerle2016a} the volume fraction $W \propto d^2$ (\textit{i.e.} the bridge cross-section) and we used this to estimate $W$ from $d$.  

We compare the modulii $E$ from the DEM simulations with experiments on similar samples (again, using $D=210$ $\mu$m beads) in Fig.~\ref{fig:Figure_E_vs_W}.  The two sets of results agree up to the end of the pendular regime (dashed line in Fig.~\ref{fig:Figure_E_vs_W}), although the simulations are again consistently stiffer than the experiments. As expected, the simulations diverge from the experiments in the funicular regime, $W>W^\star$, where the polymer in the experiments starts to form larger suprastructures, rather than simple independent bridges between particles.  In this regime, different assumptions about the elasticity of the particle bonds would be required, as only a fraction of the matrix material will be supporting the stress. 

\section{Discussion \& Conclusion}

This work has presented direct measurements of the micromechanical properties of cohesive granular media, where hard particles are connected by softer bonds.  These measurements complement previous characterisation of the bulk elasticity and failure criteria of this type of material. \cite{hemmerle2016a, schmeink2017}  A simple discrete element model,\cite{yamaguchi2020} was then used to mediate between the two limits, by upscaling the microscopic interactions in order to make predictions about larger assemblies of cohesive particles. 

Despite their minimal set of ingredients, the DEM simulations correctly predict the relationships between the microscopic nature of the bridges and the macroscopic properties of a sample, and provide insights on the underlying mechanisms.   Thus, they confirm how crucial are the details of the bond or matrix geometry to the elastic and fracture properties of cohesive granular materials. A small difference in the bridge length can significantly change, for example, the stiffness of the whole aggregate. 

In particular, the simulations reproduce two major experimental results: the scaling of the material stiffness, $E$, with the matrix stiffness, $E_p$, over several orders of magnitude; and the scaling of $E$ with the matrix content, $W$, within the pendular regime. The fact that the simulations diverge from the experimental results precisely at $W=W^\star$, which corresponds to the transition to the funicular regime, shows the role of suprastructures, such as trimers, in the softening of the samples at large~$W$. Wang \textit{et al.}\cite{wang2019} observed a similar transition in stiffness with increasing matrix content using millimetric beads and a harder cement, suggesting that a cross-over of stiffness with $W$ is a fairly general result. 

Nevertheless, the material appeared to be consistently stiffer in the simulations than experiments.  This is arguably a result of our choice of a single value for the bridge length $\delta^0$. Changing $\delta^0$ from 1 to 2 $\mu$m while calculating the bond spring constants significantly decreased the stiffness of the material, as shown in Fig.\ \ref{fig:Figure_E_vs_Ep}, and it is clear that a realistic granular material would show a distribution of values for $d$ and $\delta^0$. Indeed, a close inspection of tomography scans indicates the presence of bridges with a spanning distance significantly higher than 2 $\mu$m,\cite{schmeink2017} which would lead to a lower stiffness of the bridges and of the ensemble.

Interestingly, the model also correctly predicts the strain at which a sample ceases to have a linear response to compression and starts to fail (Fig.\ \ref{fig:Figure_comparison}), which demonstrates the accuracy of the criterion used for breaking a bond in the simulations.  Regarding Eq. \ref{criterion}, we see that bonds are more likely to break in tension than in shear, and that the breakage depends on physical parameters, such as the Young's modulus and the interfacial energy. This is a key result for future modelling of failure in such a cohesive granular medium, or for the design of this type of material.

Finally, the numerical simulations also give access to quantities difficult to measure in experimental systems, such as the spatio-temporal dynamics of microscopic failures within the sample, and a companion article\cite{yamaguchi2020} uses these signals to further explore failure in this model.    In cohesive granular materials, brittle failure is expected to be abrupt with minimal precursory damage, while a transition into a ductile mode involves significant activity prior to catastrophic breakdown, which can be analysed for precursor signals. Studies of these materials will benefit from the present relatively simple minimal model, informed in all aspects by explicit microscale quantification, which accurately predicts elastic bulk response. 

\section*{Conflicts of interest}
A.H. and L.G. are inventors on a patent regarding the cohesive granular material described in this paper (WO/2018/014935).

\section*{Acknowledgements}
We thank Matthias Schr\"oter and Ran Holtzman  for helpful discussions in the early development of this work and Soumyajyoti Biswas for generous input into the model design.  YY was supported by the Leading Graduate Course for Frontiers of Mathematical Sciences and Physics, The University of Tokyo, MEXT.  

%

\end{document}